# Analysis of Biomedical Data to Assess the Risk of Heart Rate Variability in Athletes Participating in Long-Term Excessive Endurance Exercise

Saeid Aghasoleymani Najafabadi[1,*], Hadi Jabbari Saray[2]

1. Faculty of Industrial Engineering, Urmia University of Technology, Urmia, Iran
2. Department of Computer Engineering, Urmia Islamic Azad University, Urmia, Iran

**Abstract**

Amateur marathon runners who exercise excessively over time have pathological structural changes in their hearts and aortas. Amateur marathon runners' cardiovascular system adaptations and dangers are discussed in this article. After completing endurance races, amateur athletes experience a series of cardiac modifications, including temporary elevation and changes in biomarkers of cardiac damage associated with an increased risk of coronary atherosclerosis, arrhythmias, and sudden cardiac death. As a result of the high prevalence of "false positive" biomarkers in athletes, the health benefits of aerobic activity are questioned, and treatment is complicated. Reports on long-term aerobic exercise contradict atherosclerosis risk. Differences may influence the results in lifestyle characteristics among participants. By comparing runners with their non-runner wives, we find that regular, high intensity run training improves many components of the cardiovascular profile but does not reduce atherosclerosis. Although metabolomic approaches have been developed to evaluate the physiological response of marathon runners, there is still controversy concerning the biomarkers of cardiovascular system alterations caused by long-term high-intensity endurance training. The cardiovascular risk profile is improved by habitual endurance exercise. Atherosclerosis is caused by age and cardiovascular risks; however, is little affected by it.

## 1. Introduction

Running marathons is a prime example of a major sport that places significant demands on the cardiovascular, musculoskeletal, immunological, metabolic, and nervous systems [1]. The majority of recreational marathon runners [2] throughout the last three decades have been middle-aged males. This age range is also when the atherosclerotic cardiovascular disease manifests clinically [3]. This tendency may have consequences for cardiovascular risk assessment prior to participation. A challenging aspect of marathon running is assessing at-risk runners and determining whether cardiovascular testing is necessary. It has been reported that the risk of sudden cardiac mortality linked with marathon running is too low to encourage regular coronary artery disease (CAD) screening. It is important for master athletes to have pre-participation medical exams before beginning master sports training programs. Despite coronary plaques that are susceptible to rupture, typical clinical exercise tests can be predicted. Physical activity and obesity are possible modifiable risk factors. Among the most important cardiovascular risk is the type of obesity that affects the abdomen [4]. It has been found that the size and density of epicardial adipose tissue (EAT) are correlated with intra-abdominal fat mass, the severity of obesity, and the frequency of cardiac events, respectively [5]. EAT is a visceral fat that is located between the pericardium and the heart and has a high metabolic activity [6,7].

The metabolic state is the sole factor that determines EAT activity. In people with metabolic syndrome, the perivascular tissue (per coronary adipose tissue, PCAT), a subtype of EAT, loses its protective role and develops into an aggressive, pro-inflammatory tissue that secretes cytokines and chemokines. Recent studies have demonstrated a relationship between the EAT thickness assessed by echocardiography and the carotid intima-media thickness (IMT), a strong indicator of systemic atherosclerosis, regardless of body mass index (BMI) and waist circumference (WC) [8]. Therefore, PCAT may have a direct role in the etiology of coronary artery disease [9]. Vigorous, prolonged exercise induces a large increase in the oxygen demand of skeletal muscle, as well as a significant increase in pulmonary oxygen absorption and blood transport ability to satisfy oxygen demands during skeletal muscle contraction and diastole. A number of cardiac changes are brought on by endurance exercise, including resting bradycardia, first- and second-degree atrioventricular block, increased intolerance against the upright position, thickening of the left ventricular wall, increased ventricular volume, cardiac hypertrophy, and even a slight decrease in the left ventricular ejection fraction in some endurance athletes; these adaptations occur primarily in trained endurance athletes, although the conduction abnormality is more prevalent in untrained endurance athletes. Neilan et al. [11] detected acute myocardial diastolic dysfunction in amateur middle-aged and older men after participation in the Boston Marathon. A meta-analysis of 294 amateur marathon runners found an immediate, substantial loss in left ventricular systolic function, as well as evidence of ventricular strain and distortion, after the conclusion of an event [12,13]. In general, marathon training promotes early diastolic filling of the left ventricle. Diastolic function is much better in elite marathon runners compared to non-athlete controls. Other investigations have shown that the diastolic function of the ventricle is diminished immediately after a marathon, especially in inexperienced amateur runners. Nonetheless, echocardiography has shown that these functional abnormalities return to normal after twenty-four hours following marathon training [14]. The transitory decline in systolic and diastolic ventricular function resulting from long runs is often characterized as physiological "cardiac tiredness" [14]. Further research is required to determine if amateur marathon runners may build an athlete's heart following the same path as young marathon runners.

## 2. Literature review

In order to compare the amount of energy consumed by ultra-marathon runners and sedentary individuals, Konwerski et al. [15] evaluated the relationships between EAT and coronary artery disease risk variables (CAD). Cardiovascular magnetic resonance was used to quantify the EAT volume around the right ventricle (RV), three major coronary arteries, and 30 healthy amateur ultrarunners and 9 sedentary controls. When compared to the control group, ultrarunners had a reduced percentage of plasma IL-6 that was pathologically high (17% vs. 56%) (p 0.05). IMT in both groups was comparable. EAT around the left anterior descending artery, circumflex artery, RV, and FAT percent, as well as EAT around the circumflex artery, were all positively correlated with LDL and non-HDL cholesterol in the group of ultrarunners (p 0.05 for all). Ahmadi et al. [27] integrated finite element analysis (FEA) with physics-informed neural networks (PINNs) to predict the biomechanical behavior of the human lumbar spine. Their model achieved 94.30% accuracy in estimating material properties, including Young's and shear modulus, by combining CT/MRI data with deep learning. This approach reduced manual processing and improved reliability in spine modeling. The study by Malek et al[16] included 10 controls who did not regularly participate in sports and 30 healthy, male, ultra-marathon runners (mean age 40.9 6.6 years, median 9 years of running with frequent contests). The patients underwent cardiac magnetic resonance (CMR) using a 3 T scanner, which included T1-mapping, late gadolinium enhancement (LGE), and extracellular volume (ECV) measurement. Athletes had a significantly greater left ventricular mass (LV) and larger heart chambers. The systolic function of the LV remained unaltered. According to the data, athletes with dilated cardiomyopathy or arrhythmogenic right ventricular cardiomyopathy constituted 73.3 percent of the population. Among the eight athletes and one control, 27 percent had non-ischemic, small volume LGE (p-value = 0.40). It was restricted to the septum or inferolateral wall (5 athletes and 1 control), as well as to the insertion locations (3 athletes). Maria et al. examined the effects of moderate to vigorous physical exercise on heart function, neural regulation of the cardiovascular system, and the incidence of cardiovascular diseases over a medium to long period of time [17]. At 10-year follow-up, the incidence of cardiovascular diseases was low in this group of middle-aged athletes, similar to that of the general population. The HRV study revealed a reduction in sympathetic modulation and an increase in vagal modulation toward the heart compared to B. Furthermore, active standing was associated with an increase in HRV, AP variability, and BRS indices. Last but not least, echocardiographic measurements in athletes were normal.

Aengevaeren et al.[18] recruited 12 male runners over 45 years of age for the 2017 Amsterdam Marathon. A total of six blood samples were collected: one week before the marathon, one hour before the marathon, one hour after the marathon, four hours after the marathon, one to two days after the marathon, and two weeks after the marathon (recovery). All cardiac biomarkers, with the exception of BNP, rose strongly after a marathon run, with peak values occurring 4–2 hours after the event. Farhadi Nia et al. [30] investigated the application of large language models like ChatGPT in dental diagnostics, emphasizing their potential to enhance communication, streamline workflows, and support clinical decision-making. Their study highlights how AI integration, particularly ChatGPT-4, can transform oral healthcare by providing real-time insights and assisting dental professionals. Sempere-Ruiz et al. [28] assessed the use of HRV thresholds to estimate exercise intensity in chronic heart failure patients. They found HRVT2 strongly correlated with VT2 and could serve as a valid alternative when CPET is unavailable. HRVT1 showed weaker associations, indicating limited reliability. Sundas et al. [31] reviewed five decades of HRV research, highlighting its value in assessing autonomic function and clinical conditions. They identified major gaps, including inconsistent protocols and limited clinical adoption. The study calls for standardized methods and more trials to enhance HRV's clinical utility. Gürses et al. [34] examined the effects of blood flow restricted resistance exercise (BFR-RE) on heart rate variability in young males. Their findings showed that BFR-RE had a stronger impact on cardiac autonomic function compared to traditional high-load resistance exercise. Ahmadi et al. [35] compared U-Net and Segment Anything Model (SAM) for breast tumor detection in ultrasound and mammography images. Their findings showed U-Net performed better, especially in complex cases with irregular or unclear tumor boundaries, highlighting its suitability for medical image segmentation. Zimatore et al. [38] reviewed the potential of using HRV time series and AI methods to detect cardiac fatigue early. Their findings highlight the scarcity of objective HRV-based measures and emphasize the need for improved AI-based approaches to distinguish physiological fatigue from cardiac pathology.

According to Terentes-Printzios et al. [19], 30 marathon runners were compared with 20 recreationally active controls who were age and sex matched. They used flow-mediated dilatation (FMD) of the wrist and carotid intima-media thickness (cIMT) to assess endothelial function (cIMT). There was a lower cIMT in marathon runners compared to controls and a greater FMD in marathon runners compared to controls, both of which indicate improved endothelial function in vascular blood vessels. Increasing exercise training appears to eliminate the positive impact of exercise on endothelial function. As a result of these findings, we have gained a deeper understanding of how marathon running affects cardiovascular health. Ahmadi et al. [35] conducted a comparative study of U-Net and the pretrained SAM model for tumor detection in breast ultrasound and mammography images. Their results showed that U-Net outperformed SAM in accurately segmenting tumors, especially in cases with irregular shapes and indistinct boundaries. This highlights the importance of using task-specific deep learning models for medical image segmentation. Casanova-Lizón et al. [31] evaluated HRV-based exercise in sedentary adults and found both app-guided and trainer-led methods improved fitness. The trainer-led group showed greater strength gains, but both approaches were effective. Norcéide et al. [36] proposed an integrated system using neuromorphic vision sensors (NVS) for real-time object tracking in augmented reality environments. Their approach, inspired by biological vision, enhanced power efficiency and reduced data load compared to traditional frame-based cameras. Schumacher et al. [35] used NMR-based metabolomics to study male ultramarathon runners and found prolonged metabolic changes after extreme endurance exercise. Their results offer insights for athlete monitoring and recovery strategies. Bradzionyte et al. [37] studied cardiovascular responses to exercise in adults above and below 45 years. Results showed that older adults had lower muscle oxygen saturation and higher heart rate responses, indicating greater cardiovascular strain during and after exertion.

Augustine et al.[20] examine the morphology and function of the left ventricle (LV), 24-hour central hemodynamics, and ventricular-vascular coupling in males and female marathoners and physically active people. An ambulatory blood pressure (BP) cuff was used to monitor the 24-hour central hemodynamics (BP, pulse wave velocity, PWV, wave reflection index, RIx). The ratio of arterial to ventricular elastance (Ea/Elv), which was obtained by integrating 3DE and hemodynamic data, serves as a broad indication of ventricular-vascular coupling. In comparison with male marathoners and their female untrained counterparts, women who have completed many marathons do not have lower LV function or increased aortic stiffness, and they may even possess enhanced ventricular-vascular coupling when compared to male marathoners and their female untrained counterparts. Farhadi Nia et al. [40] developed an experiential learning module focused on vestibular system models to support interdisciplinary education. The module helps students explore physiological behavior through simulations, enhancing understanding of balance mechanisms. Ousaka et al. [21] investigated the feasibility of using an electrocardiography (ECG) sensor-embedded fabric wear (SFW) during a marathon. They examined the viability of continuously collecting cardiac data using a brand new wearable with an integrated ECG sensor throughout a marathon. Despite the fact that data on 65% of runners were properly collected, the data collection was often inadequate for female runners and the early stages of the race. In order to increase the ability of the device to detect abnormal ECGs that may precede SCA, further advancements in the ergonomics and software of the device are required.

The study by Wirnitzer et al. [22] evaluated the health of endurance runners over a range of race lengths. A survey was conducted online with 245 casual runners (141 females and 104 males). In order to assess a person's health, eight factors were examined in two clusters of health-related indicators (e.g., body weight, mental health, chronic diseases and hypersensitive responses, medication use, and health-related behaviors) (e.g., smoking habits, supplement intake, food choice, healthcare utilization). Despite the null significant relationship between race distance and seven (out of eight) subhealth variables, HM runners generally had better health status than other distance runners, as indicated by domain scores of health. Although the best health was found across all race distances, endurance running provided evidence that it improved overall health and well-being. Thompson et al. [44] explored nonlinear mechanisms in bone-conducted hearing, which uses skull vibrations to stimulate the cochlea directly. Their study highlights how bone conduction enables higher-frequency hearing than traditional air conduction and discusses its potential for ultrasonic signal demodulation. Bradzionyte et al. [32] analyzed cardiovascular responses to a 6-week multimodal exercise program using advanced ECG techniques. Their findings showed significant improvements in heart rate recovery and cardiovascular function, highlighting the benefits of physical activity on cardiac health in previously inactive adults. Pūzas et al. [40] found that one year of football training significantly improved $VO_2$, stroke volume, and cardiac output in prepubertal children compared to controls. Apelland et al. [41] assessed ECG patch accuracy in endurance athletes with atrial fibrillation and found it feasible for monitoring, though recording quality varied by exercise type. In two cases, patch ECG successfully detected AF episodes confirmed by implantable monitors.

## 3. Methods and materials

### 3.1. Correlation and Risk assessment

Pearson correlation is the most popular method for analyzing numerical variables, which provides a value between 0 and 1, with 0 denoting no connection, 1 denoting a complete positive correlation, and 1 denoting a complete negative correlation. Accordingly, a correlation of 0.70 between two variables indicates a strong and favorable relationship between the variables. A positive correlation occurs when a variable rises along with variable B[23], while a negative correlation occurs when a variable rises along with variable B drops. The Pearson correlation is generally used to assess the degree of dependability between the two halves of the sample [43-45]. In order to determine the reliability of a test, the Spearman-Brown prophecy formula is applied. The Spearman–Brown approach presupposes that the test's two parts be parallel. Parallelism requires an examinee's real score to be the same across all forms, as well as a similar mean, variance, and error across all forms. In this case, Spearman-estimated Brown's full-length dependability will be greater than other metrics of internal logic[24,46]. Not all split-half estimate techniques use the Pearson correlation. Rulon offered two split-half formulations credited to John Flanagan [47]. The standard deviation of the difference in results between the two half-tests provides the basis for one formula. The equation is

$$r_w = 1 - \frac{\sigma_d^2}{\sigma_w^2} \tag{1}$$

where σw2 is the overall test variance and d is equal to Xa Xb. This formula makes two key assumptions: (a) that the variance between the two true scores for the two half-tests is constant for all test takers, and (b) that the mistakes in the two half scores are random and unrelated[24].

The other formula is

$$r_w = \frac{4\sigma_a \sigma_b r_{ab}}{\sigma_w^2} \tag{2}$$

where the score standard deviation for one test half is represented by the "σa," and the score standard deviation for the other test half is represented by the "σb." These formulae, in contrast to the Spearman-Brown formula, do not call for equal halves with equal variances. Each presupposes independently operating experimental halves. It is not necessary to use the Spearman-Brown prediction formula for either reliability estimate. Regarding the estimation of split-half dependability, Guttman made the following contribution:

$$r_w = 2\left(1 - \frac{\sigma_a^2 + \sigma_b^2}{\sigma_w^2}\right) \tag{3}$$

## 3.2. Data collection

Complicated ventricular ectopy, which includes ventricular tachycardia, is primarily encountered in populations engaging in endurance activity and is often caused by modest functional anomalies of the right ventricle and ventricular septum. Long-standing controversy exists over the severity and clinical importance of ventricular ectopy. A nasty case of ventricular ectopy might result in cardiac arrest. Marathon runners have a surprisingly low risk of abrupt cardiac death [24]. Roca et al. evaluated 79 amateur runners who competed in the 2016 Barcelona Marathon and concluded that the cardiovascular system of these runners might be influenced in different ways both before to and during the event. No research has shown that marathon running is a high-risk activity that predisposes athletes to cardiovascular system disorders. Approximately 100 master marathon runners (mean age 57.2 5.7 years, range 50–71 years) met the inclusion criteria, which were as follows: Among those ages C50, at least five have completed a full-distance marathon (42.195 kilometers) within the past three years. Heart disease experience, diabetes mellitus, angina pectoris, renal failure, musculoskeletal condition at inclusion preventing future frequent marathon running, psychiatric disorder, and unwillingness to give informed consent were among the methodological limitations. Three strategies were utilized to find participants: (1) an advertisement in a marathon magazine, (2) a press conference at the start of the study, and (3) including participants' coworkers and friends, provided inclusion criteria were met.

# 4. Results and Discussion

## 4.1. Cardiovascular risk factors

The Framingham risk score's traditional cardiovascular risk factors (CVRF) were evaluated [25]. Blood pressure measurements were taken using automatic oscillometer blood pressure equipment (Omron 705, OMRON, Mannheim, Germany). It was determined using the average of the second and third measurements, which were both taken at least three minutes apart. Using standard enzymatic methods,

HDL and LDL cholesterol [mg/dl] were tested. The definition of current smoking was defined as a recent history of smoking. The body mass index (BMI [kg/m2]) was calculated using standardized height and weight measurements.

**Table 1.** Descriptive statistics of the participants

| Metrics | Measure | N | Mean | Standard deviation |
|---|---|---|---|---|
| Age | Years | 100 | 58 | 5.8 |
| BMI | Kg/m2 | 100 | 24 | 2.1 |
| Hypertension | percent | 100 | 12 | 0.4 |
| Blood Pressure (Systolic) | mm HG | 100 | 120 | 15 |
| Blood Pressure (Diastolic) | mm HG | 100 | 75 | 9 |
| LDL | Mg/dl | 100 | 120 | 30 |
| HDL | Mg/dl | 100 | 74 | 18 |
| Diabetes | - | 100 | 0 | 0 |
| Smoke | Percent | 100 | 4.6 | - |

A systematic scientific technique called metabolomics assesses both internal and external effects of prolonged or vigorous exercise, as well as other types of training. "Sports metabolomics" refers to the use of metabolomics-related methods to the study of sports. Athlete nutrition, training monitoring, and physical performance are some of the current study priorities. In sports training, metabolomic strategies are being used more and more often, particularly in middle- and long-distance running competitions. They started by employing the subject analysis approach to analyze the collected urine. They suggested that various elements in the urine play a crucial part in the competitiveness of the athletes. The partial least-squares approach was used to further examine the urine, and it was discovered that the athletes who made it to the finals had lower amounts of certain drugs than those who did not. Additionally, individuals who did not advance to the finals in this trial had much greater urine levels of methyl nicotinamide than those who did [26].

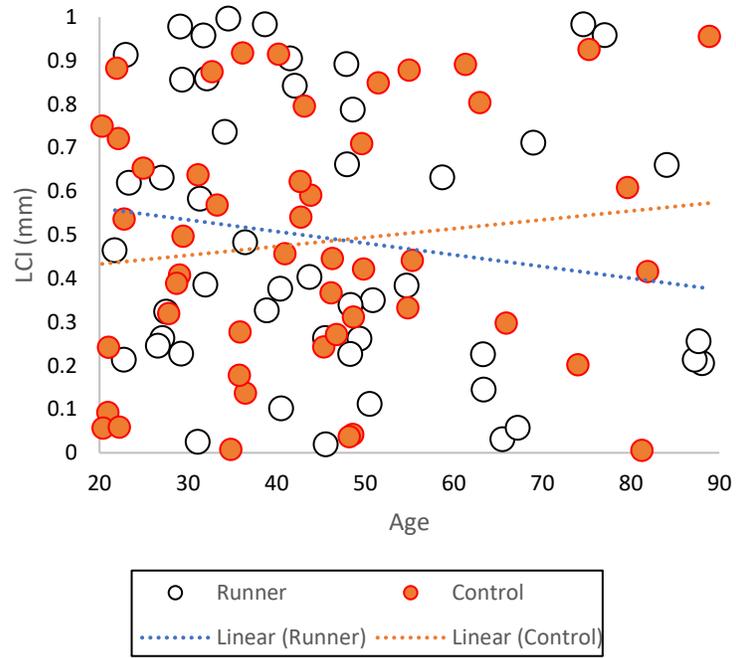

**Figure 1:** The plot of age versus left carotid

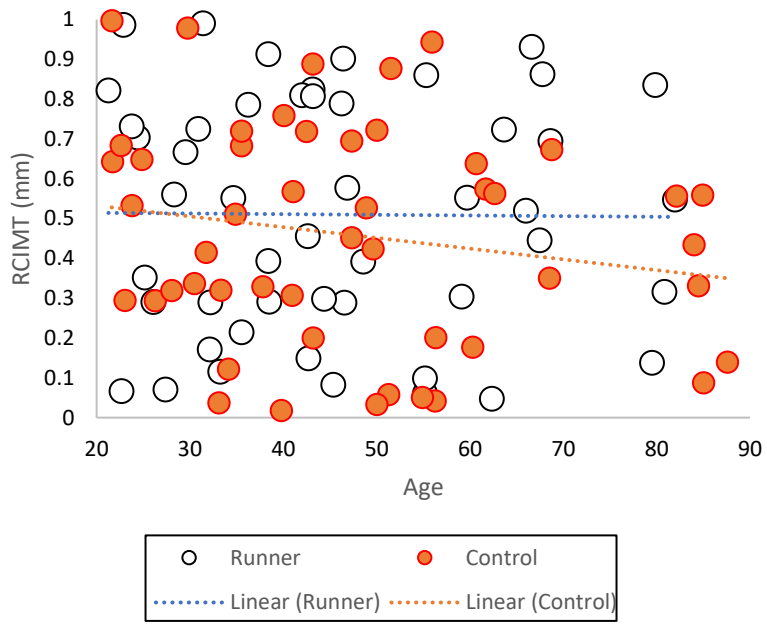

**Figure 2:** The plot of age versus right carotid

Additionally, there were no group differences in aortic SBP (p value =0.67). There was a correlation between left cIMT and aortic SBP (Pearson coefficient=0.32; p value 0.31). The connection between age and Framingham risk was r=0.41 and 0.52 (p value=0.12) for central SBP, much like it was for cIMT. Carotid enhancement tension was linked with age and the calculated Framingham risk score and did not differ across groups (p=0.67) regardless of group effects or interactions. Carotid augmentation pressure did not differ between the two groups when it was stated in proportion to a heart rate of 75 beats per minute (p=0.07) (carotid augmentation index). In a multiple linear regression model, this parameter climbed with age among groups, and it was lower in joggers. There was no correlation between the Framingham risk score and the enhancement index (all p for impacts >0.20).

Regular aerobic exercise lowers blood pressure, body weight, blood lipids, and other cardiovascular risk factors. The individual effect, however, varies greatly. In the current study, runners had body mass index that was 11% lower, CRP that was 63% lower, non-HDL cholesterol that was 13% lower, triglycerides that were 26% lower, and HDL cholesterol that was 17% higher than that of controls. In contrast, there was no difference between runners and controls in either the left or right cIMT. Similarly, marathon training had no effect on central SBP, resulting in an aging-related increase in cIMT. In light of these findings, it is reasonable to conclude that regular, intense physical activity may lower cardiovascular risk factors. Atherosclerosis, however, is neither slowed down nor accelerated by other methods such as causing vascular turbulence or affecting heart blood pressure.

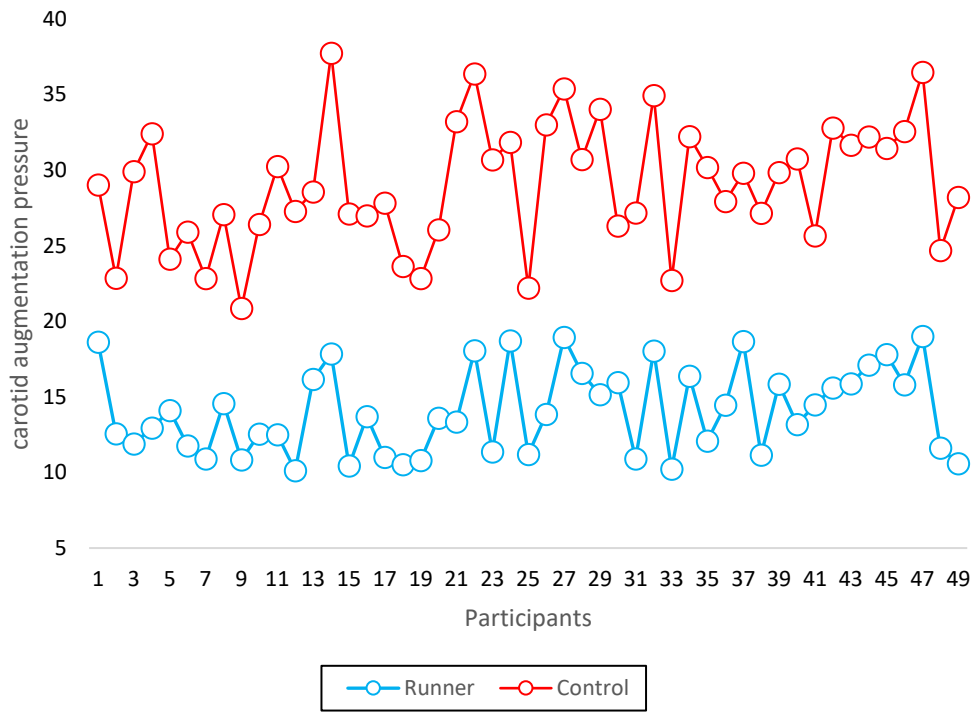

**Figure 3:** Carotid augmentation pressure for both Runner and control samples

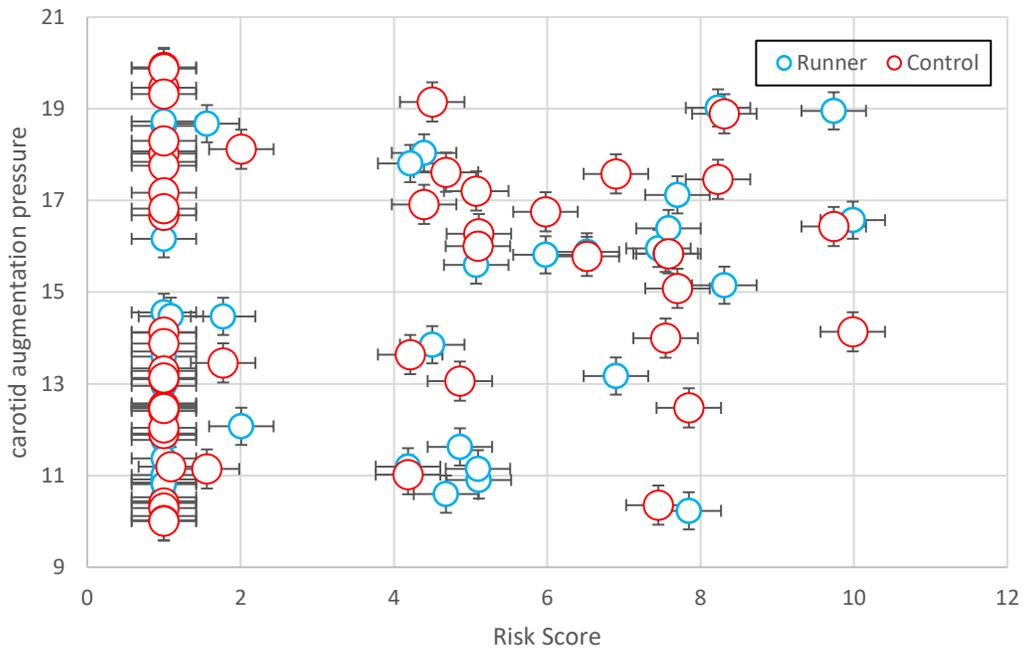

**Figure 4:** Risk factor versus Carotid augmentation pressure for both Runner and control samples

The majority of the participants started running marathons during their middle years, explaining the discrepancy between FRS and the severity of CAC in marathon runners. Therefore, exercise training might have decreased their cardiovascular risk factors. They might not accurately represent their lifetime risk exposure. More than half of our runners had smoked in the past, and 5% of them admitted to smoking now. As a result, the controls may also have been protected from the atherogenic effects of exposure to cardiovascular risk factors by virtue of their matching age and FRS. This idea has a clinical relevance in that traditional risk factor estimations may cause runners and their doctors to overestimate the athletes' actual danger. As a result of improved microvascular function, which can mask the true degree of coronary atherosclerosis and compensate for severe epicardial plaque burden, risk assessment of marathon runners becomes even more difficult. Our findings may partially explain why so many of the runners in our study were asymptomatic at rest and while exercising despite severe atherosclerosis.

## 5. Conclusion

In conclusion, there are many amateur runners, and the unacknowledged health hazards associated with marathon preparation demand our attention. The human body is an organic whole, and other systems like the immune system and homeostatic control are connected to the cardiovascular system's health. The following issues with current studies on the cardiovascular system of amateur marathon runners: 1) The majority of the subjects are middle-aged and elderly men between the ages of 40 and 55; in contrast, there are few studies on athletes between the ages of 20 and 29 who are at a high risk of sudden death in sports; 2) Domestic studies are primarily laboratory studies among professional athletes where the blood indexes are measured after long-term endurance training on test equipment, which imposes additional training and mental burdens and has a certain degree of influence;

Our findings are prone to selection bias since the recruiting methods and inclusion standards used in the two trials were different. For example, marathon runners were self-referred since the HNRS was unable to select them at random as participants. As opposed to the HNRS, they were also required to be in good health and physical condition after the age of 50 in order to qualify. Given that recurrent increases in shear stress and mechanical forces may favorably alter the calcified plaque component, it is conceivable that those who regularly engage in vigorous exercise and those who do not may have different ratios of calcified and non-calcified atherosclerotic plaque. There is currently no evidence to support this theory, which can only be examined non-invasively by using high-resolution computed tomography and extra contrast agent delivery. Our findings may help to discriminate between early subclinical cardiac abnormalities brought on by exposure to risk factors, such as arterial hypertension and aging, and physiological adaptation to exercise. Age-related declines in left ventricular end diastolic volume may contribute to marathon runners' competitiveness. A large myocardial mass may not always imply athletes' hearts in recreational runners over the age of 50, but rather a higher risk than that predicted by traditional risk factor evaluation. As a result, there is a continuing need to increase the research population, capture the topic's dynamics and breadth, find biomarkers that effectively represent sports correlations, and offer advice to amateur marathon runners on how to lower their risk of cardiovascular disease.

**Conflict of Interest**

The authors declare that the research was conducted without any commercial or financial relationships construed as a potential conflict of interest.

**The contribution of the authors**

Bahmani and Valizadeh conceived and designed the study, Adelfahmideh and Asadi collected and analyzed the data, and Akbari performed the literature search and final manuscript editing. All authors were involved in writing the manuscript. All authors read and approved the final manuscript.